\documentclass[10pt, conference]{IEEEtran}
\usepackage{booktabs}
\usepackage{multirow}
\usepackage{url}
\usepackage{cite}
\usepackage{amsmath,amssymb,amsfonts}
\usepackage{graphicx}
\usepackage{textcomp}

\IEEEoverridecommandlockouts
\IEEEpubid{\begin{minipage}{\textwidth}\ \\
\ \\
\ \\
\ \\
\copyright 2018 IEEE. DOI 10.1109/SEAA.2018.00018.
Personal use of this material is permitted. Permission from IEEE must be
obtained for all other uses, in any current or future media, including
reprinting/republishing this material for advertising or promotional purposes,
creating new collective works, for resale or redistribution to servers or
lists, or reuse of any copyrighted component of this work in other works.
\end{minipage}}

\title{Software Engineering Challenges of Deep Learning}

\begin{document}
\author{\IEEEauthorblockN{Anders Arpteg}
\IEEEauthorblockA{
  Peltarion AB\\
  Stockholm, Sweden\\
  anders@peltarion.com
}
\and
\IEEEauthorblockN{Bj{\"o}rn Brinne}
\IEEEauthorblockA{
  Peltarion AB\\
  Stockholm, Sweden\\
  bjorn@peltarion.com
}
\and
\IEEEauthorblockN{Luka Crnkovic-Friis}
\IEEEauthorblockA{
  Peltarion AB\\
  Stockholm, Sweden\\
  luka@peltarion.com
}
\and
\IEEEauthorblockN{Jan Bosch}
\IEEEauthorblockA{
  Peltarion AB\\
  Stockholm, Sweden\\
  jan@peltarion.com
}
}

\maketitle

\begin{abstract}

Surprisingly promising results have been achieved by deep learning (DL)
systems in recent years.
Many of these achievements have been reached in academic
settings, or by large technology companies with highly skilled research
groups and advanced supporting infrastructure.
For companies without large research groups or advanced infrastructure,
building high-quality production-ready systems with DL components
has proven challenging.
There is a clear lack of well-functioning tools and best practices for
building DL systems.
It is the goal of this research to identify what the main challenges are,
by applying an interpretive research approach in close collaboration with
companies of varying size and type.

A set of seven projects have been selected to describe the potential with
this new technology and to identify associated main challenges.
A set of 12 main challenges has been identified and categorized into the
three areas of development, production, and organizational challenges.
Furthermore, a mapping between the challenges and the projects is defined,
together with selected motivating descriptions of how and why the challenges
apply to specific projects.

Compared to other areas such as software engineering or database technologies,
it is clear that DL is still rather immature and in need of
further work to facilitate development of high-quality systems.
The challenges identified in this paper can be used to guide future
research by the software engineering and DL communities.
Together, we could enable a large number of companies to start taking
advantage of the high potential of the DL technology.

\end{abstract}


\section{Introduction}
\label{sec:introduction}

Deep Learning (DL) has received considerable attention in recent years due to its
success in areas such as computer vision tasks (e.g., object
recognition~\cite{krizhevsky2012imagenet} and image
generation~\cite{radford2015unsupervised}) using convolution neural networks,
natural language understanding using recurrent
neural networks (RNN)~\cite{mikolov2010recurrent} and machine strategic thinking
using deep reinforcement learning (DRL)~\cite{mnih2015human}.
One of the main differences from traditional machine learning (ML) methods is
that DL automatically learns how to represent data using multiple
layers of abstraction~\cite{bengio2000modeling, boureau2008sparse}.
In traditional ML, a significant amount of work has to be spent
on ``feature engineering'' to build this representation manually, but this
process can now be automated to a higher degree.
Having an automated and data-driven method for learning how to represent data
improves both the performance of the model and reduces requirements for manual
feature engineering work~\cite{hinton2012deep, le2013building}.

Given the recent advances in ML, we are also seeing industry
starting to increasingly take advantage of these techniques, especially in
large technology companies such as Google, Apple, and Facebook.
Google applies DL techniques to the massive amounts of data
collected in services such as Google Translator, Android's voice recognition,
Google's Street View, and their Search service~\cite{jones2014learning}.
Apple's virtual personal assistant Siri offers a variety of
services such as weather reports, sports news, and generic question-answering by
utilizing techniques such as DL~\cite{efrati2013deep}.

\subsection{Machine Learning and Software Engineering}
\label{subsec:ml_se}

Machine learning, especially DL, differs partly from
traditional software engineering (SE) in that its behavior is heavily
dependent on data from the external world.
Indeed, it is in these situations where ML becomes useful.
A key difference between ML systems and non-ML systems is that data partly
replaces code in a ML system, and a learning algorithm is used to
automatically identify patterns in the data instead of writing hard coded rules.
This suggests that data should be tested just as
thoroughly as code, but there is, currently, a lack of best practices for how to
do so.

A significant amount of research has been conducted on testing
software~\cite{kanewala2014testing}, and also for testing ML performance.
However, the intersection of SE and ML has not been so comprehensively
studied~\cite{breck2016test}.
It is not only the correctness of the model that needs
testing, but also the implementation of a production-ready ML
system~\cite{murphy2007approach}.

\IEEEpubidadjcol
\subsection{Big Data and Deep Learning}
\label{subsec:big_data}

When DL is combined with the recent growth of big data,
opportunities and transformative potentials exist for sectors such as
finance, healthcare, manufacturing, and educational
services~\cite{manyika2011big, lin2012large, smola2010architecture, chu2007map,
panda2009planet}.

Working with big data adds extra requirements to supporting infrastructure.
Instead of simply loading all data in memory on a single machine, special
infrastructure may be needed such as Apache Spark~\cite{zaharia2012resilient},
Apache Flink~\cite{carbone2015apache}, or Google
DataFlow~\cite{akidau2015dataflow}.
Big data is often loosely defined by the three Vs: Volume, Variety, and
Velocity~\cite{beyer2012importance}.
It is not only the volume of data that may require big data techniques, but
also the variety of different types and formats of data.
This is another strength of DL, which can use representation
learning to combine different modalities such as images, audio, text, and
tabular data~\cite{ngiam2011multimodal,kaiser2017one}.
However, that also presents challenges in how to integrate and transform
multiple data sources~\cite{chen2014big}.
Given the large number of different formats for images, video, text, and
tabular data, it can be challenging to build a data integration pipeline.
Additionally, velocity requirements (i.e. processing time or real-time
low latency demands) may also require the use of big data techniques such as
streaming processing.

This adds numerous challenges to the task of building a big data DL
system.
It is not only the process of extraction, transforming and loading (ETL)
data but also that novel distributed training algorithms may need to be used.
Especially DL techniques are not trivially parallelized and again
need special supporting infrastructure~\cite{dean2012large,yahoo2017tensor}.

\subsection{Technical Debt}

Technical debt, a metaphor introduced by
Ward Cunningham in 1992, can be used to account for long-term costs incurred
by moving quickly in SE.
It has been argued that ML systems have a special capacity for incurring
technical debt~\cite{sculley2015hidden}.
ML systems not only experience code-level debts but also dependencies
related to changes to the external world.

Data dependencies have been found to build similar debt as code dependencies.
Few tools currently exist for evaluating and analyzing data dependencies,
especially when compared to static analysis of code using compilers and build
systems.
One example of a data dependency tool is described in~\cite{brendan2013ad},
where data sources and features can be annotated.

Deep Learning also makes it possible to compose complex models from a set of
sub models and potentially reuse pre-trained parameters with
so called ``transfer learning'' techniques.
This not only adds additional dependencies on the data, but also on external
models that may be trained separately and may also change in configuration
over time.
External models add additional dependency debt and cost of ownership for ML
systems.
Dependency debt is noted as one of the key contributors to technical
debt in SE~\cite{morgenthaler2012searching}.

It is not uncommon that the supporting code and infrastructure incur significant
technical debt, and care should be taken to properly plan for time needed to
develop supporting plumbing code and infrastructure.

A mature system might end up with 95\% of the code being plumbing and
glue code, which mainly connects different ML libraries and packages
together~\cite{sculley2015hidden}.
An ML system is usually dependent on many different pipelines that may also be
implemented in different languages, formats, and infrastructure systems.
It is not uncommon for pipelines to change, add and remove fields, or become
deprecated.
Keeping a deployed production-ready ML system up to date with all these changes
require a significant amount of work, and requires supporting monitoring and
logging systems to be able to detect when these changes occur.

Another common bad practice in ML systems is to have experimental code paths.
This is similar to \emph{dead flags} in traditional
software~\cite{morgenthaler2012searching}.
Over time, these accumulated code paths create a growing debt that will be
difficult to maintain~\cite{sculley2015hidden}.
An infamous example of experimental code paths was Knight Capital's system
losing \$465 million in 45 minutes, apparently because of unexpected behavior
from obsolete experimental code paths~\cite{securities2013sec}.

Another technical debt problem of ML systems is configuration management.
Configuration of ML systems is frequently not given the same level of quality
control as software code, even though it can both have an higher impact on the
performance of the model and also be large in size~\cite{zheng2014challenges}.
Configuration changes should be reviewed and possibly even tested in the same
way as code changes.

The contribution of this paper is threefold.
First, we present and describe seven example ML systems for varying use cases
and we identify the potential of ML and specifically DL as a technology.
Second, we identify the key SE challenges associated with building DL systems.
Finally, we validate the relevance of these SE challenges by
providing an overview of the example systems in which we have experienced
associated challenges.

\section{Research Approach}
\label{sec:research_approach}

The findings in this paper are based on seven example ML projects that have been
carried out in close collaboration with companies of different sizes and types.
The companies range from start-up size to large multinational companies with
thousands of employees and many millions of active users.
The examples are focused on ML projects of different types where some systems
have been in production for more than 10 years, starting from 2005 and onward,
and other systems are still in the prototype stage.
Experience is collected from people directly employed in associated
organizations and working with given case studies, taking a participant
observer approach and making use of well-established case study
methods~\cite{eisenhardt1989building}.

The focus of this study is limited to identifying challenges specifically related
to the intersection of SE practices and DL applications.
The challenges are based on existing research and validated by empirical studies
of selected example projects, and the study adopts an interpretive research
approach~\cite{walsham1995interpretive}.
Software development is seen as an activity conducted by people in organizations
with different values, different expectations, and with different strategies.
This type of case study research is appropriate to explore real-life challenges
where control over the context is not possible and where there is an interest
in accessing people's experiences related to case studies and
associated organizational structures~\cite{walsham1995interpretive}.

\section{Real-World ML Projects}
\label{sec:real_world_projects}

A diverse set of real-world ML projects has been selected for this research and
are described in this section.
The projects have been selected to collectively represent and exemplify
different aspects of challenges.
A mapping between selected projects and challenges is presented in a later
section.

\subsection{Project EST: Real Estate Valuation}
\label{subsec:proj_est}

Real estate valuation is a cornerstone of financial security for banks.
When issuing loans, they need to make sure that the sales price is reasonably
close to the market value and they need to know the total value of their
collateral securities.
Project EST is a long-running (\textgreater 10 years) neural network
based real estate valuation system used in production by multiple Swedish banks.

The data used by the network consist of historical sales, information about
individual properties, demographics, and geography.
It was developed by a small team of two data scientists and two backend
developers.
It initially made use of data from a pre-existing SQL database system and a
large collection of legacy data pipeline scripts running on assorted systems.

\subsection{Project OIL: Predicting Oil and Gas Recovery Potential}
\label{subsec:proj_oil}

When building multi-stage hydraulically fractured horizontal wells in tight
gas and oil reservoirs, the main question of importance is how much oil and
gas will ultimately be recovered.
Many different methods have been proposed and used to obtain an estimate of the
Estimated Ultimate Recovery (EUR).
Project OIL, started by a US oil and gas exploration company, investigated the
use of DL in predicting the EUR based on geological information.

The data consisted of high resolution geological maps combined with a small
number of oil/gas wells.
These wells have been in production for a sufficiently long period of time to
accurately be able to deduce EUR.
The project resulted in a decision-support tool that given coordinates in the
Eagle Ford Shale could predict the EUR of an average multi-stage hydraulically
fractured horizontal well.
The development team consisted of two senior data scientists/developers.

\subsection{Project RET: Predicting User Retention}
\label{subsec:proj_ret}

An important metric for any end-user facing company is user retention, e.g.
what percentage of users will remain active two weeks after registration.
Trying to estimate the percentage of users who will remain active shortly
after registration is an important but difficult problem given the small amount
of available data per user.
Project RET refers to a set of projects built by different teams for this
purpose for a media-streaming service.
Even though the service may have many millions of active users,
being able to predict second-week retention given only few days of data per
user is non-trivial.

A number of data sources were used, such as media consumption,
client interactions (e.g. mobile application actions), and user demographics.
Assuming there are many millions of active users and that data has to
be extracted from multiple data sources, this becomes a big data problem with
many terabytes to process.
Multiple teams are working with this problem, often with a specific
functionality of the service in mind.
Each team usually consists of 4--10 data scientists and engineers.
Predictions are usually tracked on dashboards and can be used to evaluate,
for example, randomized tests of product changes.

\subsection{Project WEA: Weather Forecasting}
\label{subsec:proj_wea}

Weather forecasts today rely on physical modelling and are primarily solved
using finite element methods.
They require expensive supercomputers, take a long time to perform a single
prediction and, for many applications, are rather inaccurate.
Project WEA, performed in collaboration with a national meteorological agency,
aimed to use DL to do weather forecasts.
The ultimate purpose was building better wind turbine generator predictions.

The data, which was large (\textgreater 1 TB), consisted of satellite images,
meteorological observations and topological maps.
The data spanned a period of 12 years.
Some of it was easily obtainable while other parts required the physical
transport of data tapes.
The project team consisted of three data scientists, two developers and a
number of meteorologists.

\subsection{Project CCF: Credit Card Fraud Detection}
\label{subsec:proj_ccf}

For companies in the gaming industry, credit card fraud can be a significant
issue due to the possibility of money laundering.
If money is deposited using a stolen credit card, it is possible withdraw
``winnings'' from another account after losing deliberately or
after turning the money round a few times in casino-style games.
Preventing or detecting fraudulent deposits early on by combining payment
details with activities in the games is crucial.

Project CCF was started with the purpose of building a generalized model for
this purpose.
Significant time was spent on improving the model efficiency of fraud detection
and, in so doing, also reducing levels of blocked, risky payments in a company
where the detection was based on hand-crafted rule sets.
The data consisted of payment request data (method, amount location, etc.)
and customer details (CRM, payment history, activity history).
If the payment was approved, subsequent analysis also took into account
post-payment activity such as signs of abnormal play.
A small dedicated analysis team was responsible for developing the models, and
results were handed over to a different department for integration into the
payment processing systems.

\subsection{Project BOT: Poker Bot Identification}
\label{subsec:proj_bot}

Online poker grew very quickly in the early 2000s.
With a large number of inexperienced players, waging real money there was an
opportunity to make substantiate winnings for experienced players.
However, the stakes were relatively low on the tables favored by inexperienced
players so a large number of games had to be played to make a reasonable profit.
Therefore, many attempts were made (some successfully) to automate game-playing
by writing software, i.e. ``poker bots''.
For the game sites, it was and is a question of customers' trust to keep the
games free from these bots.
Project BOT was initiated to develop methods for detecting automated
game-playing, and to be able to quickly lock associated accounts.

The data used was game play activity (i.e. statistics on actions taken given
the state of the game as seen from each player's perspective) together with
client technical details such as connection IP, hardware fingerprints, player
clicking/timing statistics.
The models were developed by a small consultancy team (2--3 people) together
with internal experts (1--2 people).
Despite promising results, the project was cancelled before it was finished due
to unpredictability of efficiency of the finished product.

\subsection{Project REC: Media Recommendations}
\label{subsec:proj_rec}

The field of recommendation systems has increased in popularity and
importance with the uprise of media-streaming services.
Having a high-quality recommendation service that does not only understand the
users' preference but also specific user contexts can significantly improve the
user experience.
Traditional techniques such as collaborative filtering work well given that
sufficient data exist for both the user and the content.
However, it can become problematic when new content is released
such as a new movie or song that few users have seen.
Being able to quickly recommend new content is important, as users are
interested in novel content.

Project REC refers to the work with solving the problem of being able
to recommend new content.
Using techniques such as deep convolutional networks, it
is possible to use e.g. raw media data to learn user preferences for novel
content without having any user usage data.
Having a hybrid system that makes use of collaborative filtering techniques
when sufficient user data exist, and DL for novel content, offers
the best of two worlds.
This DL project started as a small prototype project, which was later expanded
into a full production-ready system by a team of data scientists and a
large number of engineers.

\section{Selected Challenges}
\label{sec:challenges}

This section presents a list of concisely described challenges in the
intersection between ML and SE.
They have been grouped into three categories: development, production,
and organizational challenges.

\subsection{Development Challenges}
\label{subsec:dev_challenges}

There are fundamental differences between developing ML systems compared to
traditional SE systems.
One of the main differences is that data is used to
program the system ``automatically'' instead of writing the software code
manually.
The performance of the system is unknown until it has been tested with
given data, making it difficult to plan the project in a structured manner.

In addition, the lack of model transparency, inability to understand large and
complicated models, and difficulty in debugging using libraries with lazy
execution makes it challenging to estimate the effort needed to complete the
project.

\subsubsection{Experiment Management}
\label{subsubsec:exp_management}

During the development of ML models, a large number of experiments are usually
performed to identify the optimal model.
Each experiment can differ from other experiments in a number of ways and it
is important to ensure reproducible results for these experiments.
To have reproducible results, it may be necessary to know the exact version of
components such as:

\begin{enumerate}
    \item Hardware (e.g. GPU models primarily)
    \item Platform (e.g. operating system and installed packages)
    \item Source code (e.g. model training and pre-processing)
    \item Configuration (e.g. model configuration and pre-processing settings)
    \item Training data (e.g. input signals and target values)
    \item Model state (e.g. versions of trained models).
\end{enumerate}

Version control for ML systems adds a number of challenges compared to
traditional software development, especially given the high
level of data dependency in ML systems.
Different versions of data will yield different results, and the input data
are often a conglomerate of data from multiple heterogeneous data sources.
It has been argued that one of the most difficult components to keep track of
is the data component, and the cost and storing of versioned data can be very
high~\cite{morgenthaler2012searching}.

Furthermore, different versions of the model are created during training of the
DL model, each with different parameters and metrics that need to
be properly measured and tracked.
With the addition of data dependencies and a high degree of configuration
parameters, it can be very challenging to properly maintain ML systems in the
long run.
Also, it is not uncommon to perform hyperparameter tuning of models,
potentially by making use of automated meta-optimization methods that
generate hundreds of versions of the same data and model but with
different configuration parameters~\cite{golovin2017google}.
Deep learning can also add the requirement of specific hardware.

\subsubsection{Limited Transparency}
\label{subsubsec:lim_transparency}

Software engineering is based on the principle of reducing complex systems
into smaller, simpler blocks.
Whenever possible, it is desirable to group blocks into different levels of
abstraction that have a similar conceptual meaning.
Although DL systems, in principle, do that automatically, it is
very difficult to know exactly how it is performed or predict what the
abstraction layers will look like once the model has been trained.
Furthermore, it is difficult to isolate a specific functional
area or obtain a semantic understanding of the model.
This can only be performed with approximated methods~\cite{bengio2012deep}.

The great advances that have been made in fields such as computer vision and
speech recognition, have been accomplished by replacing a modular processing
pipeline with large neural networks that are trained
end-to-end~\cite{lecun2015deep}.
In essence, transparency is traded for accuracy.
This is an unavoidable reality.
If the problem was simple enough to be explained in symbolic logic, then there
would be no need for complex system to solve it.
However, if the problem is complex and the model is inherently irreducible,
then an accurate explanation of the model will be as complex as the model
itself~\cite{glorot2010understanding}.

\subsubsection{Troubleshooting}
\label{subsubsec:troubleshooting}

A major challenge in developing DL systems is the difficulties in
estimating the results before a system has been trained and tested.
Furthermore, our poor understanding of the inner workings of complex
neural networks makes it difficult to debug them in a traditional way.
In a neural network, the structure combines the functional parts and the
memory and can also be distributed across multiple machines.

In addition, using libraries such as TensorFlow~\cite{abadi2016tensorflow}
potentially combined with big data frameworks such as Apache
Spark~\cite{zaharia2012resilient} makes it difficult to troubleshoot
and debug problems in the code.
As they both have a lazy execution graph, where the code
is not executed in imperative order, it can be difficult
to troubleshoot bugs using traditional SE tools.
Other frameworks such as PyTorch~\cite{pytorch2017pytorch}, do not have the
lazy execution graph problem but have other issues such as lower adoption rate
in the AI community.

Even if it was possible to step through the source code and set break points,
manual evaluation is, in practice, often impossible as it would involve the
inspection of millions of parameters.
Compared to traditional SE, a small bug in the source code
may not be detected at neither compile- nor run-time.
During training, the only information that the developer or data scientist
may have is a global error estimate.
As large neural networks sometimes take days or even weeks to train,
there is no way of guaranteeing that a certain performance level will ever
be reached.
Hence, the effort required to build them can become very
large~\cite{chen2014big}.

\subsubsection{Resource Limitations}
\label{subsubsec:resource_lim}

Working with data that require distributed system adds another magnitude of
complexity compared to single machine solutions.
It is not only the volume of the data that may require a distributed solution,
but also computational needs for extracting and transforming data, training
and evaluating the model, and/or serving the model in production.
For DL systems, it can also be limited GPU memory that requires
special techniques to split the model across multiple GPUs.

Working with distributed systems, both for data processing such as Apache
Spark~\cite{zaharia2012resilient} and DL training such as
Distributed TensorFlow or
TensorFlowOnSpark~\cite{abadi2016tensorflow,yahoo2017tensor}, adds complexity in a number of
dimensions.
It not only requires additional knowledge and time to operate them, but also
additional management and cost of associated hardware and software.

\subsubsection{Testing}

ML systems require testing of software used for building
data pipelines, training models, and serving in production.
Given the high data dependency of ML systems, data also need to be tested.
However, currently only a few data-testing tools exist compared to software
testing.
A frequent pattern seen when testing data is to make use of a small sample
of the full dataset.
It is challenging to provide a sample that includes all the edge cases that
may exist in the full dataset.
Also, as the external world is dynamic and changes over time, new edge cases
will continue to appear later in time.

In addition, the non-deterministic nature of many training algorithms makes
testing of models even more challenging.
Another challenge can be that data processing and model serving in production
mode can differ in implementation compared to training or testing mode,
causing a training-serving skew in model performance~\cite{tata2017quick}.
Having the proper tests in place then becomes crucial.

\subsection{Production Challenges}
\label{subsec:production_challenges}

Particularly for DL, it is important to take advantage of the latest
hardware.
This yields a significant challenge to maintain frequent updates to
state-of-the-art software and managing associated dependencies.
To accurately detect problems introduced by changing behaviour in
dependencies, including data sources that have been modified, requires
careful and clever monitoring.

More interestingly, since ML systems are often used directly by end-users,
the model might cause a change in the reality it tries to understand.
Hence, it may introduce a hidden feedback loop between the production and
training data used by the model.

\subsubsection{Dependency Management}
\label{subsubsec:dependency_mgm}

Traditional SE typically builds on the assumption that
hardware is at best a non-issue and at worst a static entity that has to be
taken into consideration.
DL systems are primarily trained on GPUs as they provide a 40-100x
speedup over classic CPUs.
For the past 5 years, the performance has significantly improved and new
GPUs are released 1--2 times per year.

As each step increases the performance significantly (beyond Moore's
Law)~\cite{geer2005taking} and hardware performance directly translates to reduced
training time (or better results for equivalent training time) there is a great
incentive for the software to tightly follow the hardware development.

The DL software platforms are continuously updated on a weekly and
sometimes daily basis and the updates typically result in noticeable
improvements.
This works well for academic research and for developing proofs
of concept but can cause considerable issues for production-ready
systems.
Unlike other ML methods, DL often scales
directly with model size and data amount.
As training times can be very long (typically a few days up to several weeks)
there is a very strong motivation to maximize
performance by using the latest software and hardware.

Changing the hardware and software may not only cause
issues with being able to maintain reproducible results, they may also incur
significant engineering costs with keeping software and hardware up to date.

\subsubsection{Monitoring and Logging}
\label{subsubsec:monitoring}

Building a toy example ML system or even an offline research prototype is
easy compared to the amount of work required to build a production-ready ML
system~\cite{breck2016test}.
In real-world ML applications beyond toy examples, it can become
difficult to cover all the edge cases that may occur once a model has been
deployed in production.
It is also common that people fail to recognize the effort needed to maintain a
deployed ML system over time~\cite{sculley2015hidden}.

An ML system may be retrained frequently and thus change behavior autonomously.
As the behavior of the external world changes, the behavior of the ML system
can suddenly change without any human action in ``control'' of the system.
In this situation, unit testing and integration tests are valuable
but not sufficient to validate the performance of the system.
Old thresholds that may have been manually assigned may no longer be valid
given drifts in the data from the external world.

Live monitoring of the system performance can help, but choosing what metrics
to monitor can be challenging.

\subsubsection{Unintended Feedback Loops}
\label{subsubsec:feedback_loops}

A ML system is, by definition, always open-ended as it is driven
by external data.
No matter how carefully the model is designed and tested, its final performance
will always be heavily dependent on the external data~\cite{sculley2015hidden}.

Furthermore, especially in models deployed in a big data context
(as ML systems often are), there is a risk of creating an unintended
feedback loop where real-world systems adapt to the model
rather than the other way around.
Imagine having a widely used real estate price prediction system.
When such a system becomes sufficiently popular,
the predictions can easily become a self-fulfilling prophecy.
As known from control theory, apart from controlled special cases, positive
feedback loops are inherently unstable~\cite{bottou2013counterfactual}.

\subsubsection{Glue Code and Supporting Systems}
\label{subsubsec:glue_code}

An unfortunate property of ML systems, and especially DL systems,
is that only a small part of the system deals with the model.
In a production-ready system, only 5\% of the code may deal with the model
and the rest is ``glue code'' that interacts with supporting systems and glues
libraries and systems together~\cite{sculley2015hidden}.

Using cloud services can greatly improve development time and decrease
maintenance needs.
However, keeping the glue code up to date, and keeping up with external changes
in cloud services, can introduce unexpected challenges in production-ready
systems.

\subsection{Organizational Challenges}
\label{subsec:org_challanges}

To put an ML model into production usually requires collaboration between
many different teams with different ideas, priorities, and cultural values.
This not only introduces organizational challenges from a cultural
point of view, but also in being able to properly estimate the amount of
effort needed by the different types of teams.

\subsubsection{Effort Estimation}
\label{subsubsec:effort}

The reductionist modular SE design of a non-ML project makes it considerably
easier to estimate the time and resources required to complete it.
In a ML project, the goal might also be well defined but it is unclear to
what extent a learned model will achieve that goal, and an unknown number of
iterations will be needed before the results reach acceptable levels.
This is in the nature of any research project.
It is can also be difficult to decrease scope and run the project in a
time-based setting, with a predefined delivery date, since it is hard to
determine when, if at all, acceptable performance will be achieved.

An added complication is the lack of transparency inherent in many ML
models, especially in DL with powerful but complex and poorly understood
models~\cite{glorot2010understanding}.
Since, in many cases, there is no easy way to understand how a model works,
it is also very difficult to understand how it should be modified to reach
better results.

\begin{table*}
\centering
\caption{Challenges faced in selected projects}
\label{tab:mappings}
\begin{tabular}{l|l|ccccccc}
\toprule
Category & Challenge            & EST & OIL & RET & WEA & CCF & BOT & REC\\
\midrule
\multirow{5}{*}{Dev}
& Experiment Management &  X  &  X  &  X  &  X  &     &  X  &  X \\
& Limited Transparency  &  X  &  X  &     &  X  &  X  &     &  X \\
& Troubleshooting       &  X  &     &  X  &  X  &     &     &  X \\
& Resource Limitations  &  X  &     &  X  &  X  &     &     &  X \\
& Testing               &  X  &  X  &  X  &  X  &  X  &  X  &  X \\
\hline
\multirow{4}{*}{Prod}
& Dependency Management         &  X  &     &  X  &  X  &  X  &     &  X \\
& Monitoring and Logging        &     &     &  X  &  -  &  -  &  -  &    \\
& Unintended Feedback Loops     &  X  &     &     &     &     &     &  X \\
& Glue Code and Supporting Systems &  X  &     &  X  &  X  &     &     &  X \\
\hline
\multirow{3}{*}{Org}
& Effort Estimation     &  X  &     &  X  &     &  X  &  X  &  X \\
& Privacy and Safety    &  X  &     &  X  &     &  X  &  X  &    \\
& Cultural Differences  &  X  &  X  &  X  &     &  X  &  X  &    \\
\bottomrule
\multicolumn{2}{l}{} &
\multicolumn{7}{l}{``X'': the project has clearly experienced associated challenge} \\
\multicolumn{2}{l}{} &
\multicolumn{7}{l}{``-'': the challenge is not applicable to the associated project}
\end{tabular}
\end{table*}

\subsubsection{Privacy and Data Safety}
\label{subsubsec:privacy}

A lack of understanding of the internal workings of a large neural network
can have serious implications for privacy and data safety.
The knowledge in a neural network is stored in a distributed manner across the
weights of the network.
Although we know that specialization occurs in specific regions of the neural
network, its exact mechanism is poorly understood.
Thus, it is very difficult for designers to control where and how information
is stored.
It also not uncommon for companies to have terms of service agreements
with their end-users that prevents them from using raw data as direct input
to a ML model, and instead have to make use of anonymized and/or
aggregated statistics of the user data~\cite{tata2017quick}.
This can not only reduce the performance on the model, but can also make tasks
such as data exploration and troubleshooting problems more difficult.
New regulations such as the European General Data Protection
Regulation~\cite{hustinx2017eu} go to great lengths to keep the data safe and
to protect privacy concerns.
However, while keeping data safe is important, it also adds a significant
challenges in how to develop and manage ML systems.

Although the information in a model is obscured and there is no trivial way of
transforming it back to humanly readable information, it is not impossible to
do so.
There has been some work on preserving the safety and privacy of data,
e.g.\ differential privacy~\cite{dwork2008differential},
k-anonymity~\cite{sweeney2002anonymity}, and encrypted networks that make use
of homomorphic encryption to keep sensitive data safe~\cite{xie2014crypto}.

However, more work is needed to preserve the privacy and safety of sensitive
datasets while still being able to efficiently perform data exploration,
develop models, and troubleshoot problems.

\subsubsection{Cultural Differences}
\label{subsubsec:cultural}

Building a production-ready ML system usually involves a collaboration between
people with different roles.
A data scientist might be ``pragmatic'' about their code
as long as it achieves the desired results in a controlled environment, whereas
members of the engineering teams care much more about maintainability and
stability.
Transforming an initial prototype into a production-ready system that
also interacts with existing backend and frontend systems usually requires
a significantly larger effort.
This normally includes collaboration with, for example, backend engineers,
UX designers, and product owners.

It is not uncommon that the culture, skills, and interest areas differs
between these people.
For example, data scientists may be lacking in SE
skills and understanding of good engineering practices.
UX designers, who have a good understanding of how to optimize the
user experience, may also have a different culture and ways of working that
can introduce challenges in working together to develop a production-ready
ML system~\cite{tata2017quick}.

\section{Project Challenge Mappings}
\label{sec:mapping}

Given selected projects and identified challenges, a mapping between the two
is provided in Table ~\ref{tab:mappings}.
The top row of the table lists the three letter acronyms for respective
project, e.g. EST refers to ``Project EST: Real Estate Valuation''.
An ``X'' means that the project has clearly experienced the specific challenge.
A ``-'' (dash) means that the challenge is not applicable for the specific
project, e.g. because the project was never deployed.
To further motivate and explain the mappings, see respective motivating text
below.
Due to the limited space in this paper, we are unable to describe
every mapping in the table, but motivating highlights for how projects are
related to challenges are provided.

\subsection{Experiment Management}
\label{subsec:exp_mgm}

In the Real Estate Valuation (EST) project, a large number of
experiments were conducted, and the code behind pre-processing, training, and
evaluation of the models were continuously improved.
During one code refactoring in this project, a data shuffle flag was
accidentally switched from false to true.
This caused a significant decrease of the model performance.
Several days of work were lost in identifying and resolving this shuffle problem.
This illustrates one of the challenges with understanding differences between
experiments, merely understanding potential code modifications between
experiments is hard, especially whether or not they influenced the model
performance.
Being able to efficiently compare software, platform environment,
configurations, and potential changes in the data is a difficult but
important challenge.

\subsection{Limited Transparency}
\label{subsec:lim_transparency}

In the OIL project, a number of very expensive engineering decisions had to
be made based on the results.
Since the system was not perfect and the data could be noisy, the engineers
responsible for the exploration wanted to know why the neural network made a
certain decision.
Given that it is virtually a black box model, no direct simple explanation could
be made.
Sensitivity analysis was used to estimate the impact of the various geological
parameters, but this could not take into account when data were locally of poor
quality.
Subsequently, the engineers were highly reluctant to risk drilling a
bore hole (cost \textgreater \$ 1M) without additional corroboration.

\subsection{Troubleshooting}
\label{subsec:troubleshooting}

One of the principle components of the weather prediction system (WEA) was an
autoencoder that compressed weather data.
Early in the testing, the resolution of the reconstructed output was of very
poor quality.
There was no immediately obvious explanation and several hundreds of
experiments with different neural network models was done over a period of
two weeks.

Because of the lack of easy to use debugging tools for deep neural networks,
it took a long time to identify the cause of the error.
It turned out to be a pooling (subsampling) operation that was too aggressive,
leading to a loss of resolution before the data were actually encoded.

\subsection{Resource Limitations}
\label{subsec:resource_lim}

There can be many reasons for the need of special techniques to handle resource
limitations such as a lack of memory (CPU or GPU), long training
time, or low-latency serving needs.
One common reason is the need for distributed data processing,
to pre-process and prepare the data before training the model.
For example, the media recommendations service has both a collaborative
filtering part and a DL part.
The DL part learns how to recommend novel media that has only been seen
a few times by a user, and it needs to process raw media data.
This involves processing data at petabyte scale, and also training a rather
computationally expensive convolutional neural network.

These resource demands are extensive, involving many days' processing on
hundreds of machines.
This is an example of having both a big data processing challenge and
distributed DL training problem.

\subsection{Effort Estimation}
\label{subsec:effort_est}

The difficulties in estimating sufficient time and resources were clearly
experienced during the development of Project BOT. The business owners grew
increasingly impatient and decided to cancel it even though it had proceeded
quite well according to the ML team.

Until a working model is designed, it is very hard to claim that anything of
value has been accomplished to the client.
Thus, being unable to set a final delivery date may lead to projects being shut
down despite promising intermediate results.

\subsection{Privacy and Safety}
\label{subsec:privacy}

An example of how efforts to ensure the privacy can cause challenges can be
seen in the User Retention project.
As part of the terms of service agreement with the users of the service,
personal data needs to be anonymized and encrypted. This includes any
information that might be used to reverse-engineer the identify of the user,
e.g.\ gender, age, and city.

Even though this information would be useful to have and could improve the
performance of the model, it is encrypted and can be difficult to make use of
when training models and building the system.

\subsection{Cultural Differences and Logging}
\label{subsec:cult_diff}

As mentioned in the description of the challenge ~\ref{subsubsec:cultural},
there are many examples of how people with different skills and cultures
need to interact.
One example of cultural differences and difficulties interacting between
data scientists and product owners was experienced in the User Retention
project.
One purpose for this project was to be able to predict second-week user
retention, which partly can be used to evaluate user experience in randomized
tests (A/B tests).

A product owner has many responsibilities, including being able to push changes
to production in time.
As it may take time to, for instance, implement sufficient logging and to analyze the
results, a product owner may be unwilling to spend sufficient time
implementing necessary logging.
Without proper logging in place, there will not be sufficient data to make
accurate retention predictions.
This can result in discussions between data scientists and
product owners, that may not only have different goals, but also different
mindsets in how decisions should be made.

\section{Related Work}
\label{sec:related_work}

A number of dimensions have been explored to understand the challenges in ML
systems.
Aspects and risk factors of hidden technical debt for ML systems
were explored in~\cite{sculley2015hidden}, including risk factors such as
boundary erosion, entanglement, hidden feedback loops, undeclared consumers,
data dependencies, configuration issues, and changes in the external world.

Traditional SE practices have shown the value of clear
abstraction boundaries using, for example, encapsulation and modular design.
However, ML systems with complex models can erode these boundaries, partly
due to the dependency to the external world.
ML systems also tend to entangle signals and parameters with each other.
The CACE principle states that Changing Anything Changes Everything.
A model is not only sensitive to changes in data from the external world,
but also, for example, hyperparameters for the model such as number of layers,
learning rate, and regularization parameters.

With the addition of transfer learning where complex models are composed of
sub models, this can lead to ``correction cascades'' where
a small improvement in a dependency may lead to a decrease in performance
of the consuming system.

The problem of having undeclared consumers of ML systems, where consumers
silently use the output of the system, is similar to what is referred to
as ``visibility debt'' in traditional SE~\cite{morgenthaler2012searching}.
This tight undeclared coupling of systems can lead to use in ways that are
intended and poorly understood.
Additionally, these undeclared consumers can introduce hidden feedback loops
between the systems.

A scoring system for production-readiness of ML systems was proposed
in~\cite{breck2016test}.
Scoring 0 points means it is more of a research prototype
rather than a production-ready system.
Scoring 12+ points means there are exceptional levels of automated testing and
monitoring.
They focus on the ML specific tests needed for a production-ready
system such as tests for: 1) features and data, 2) model development, 3)
infrastructure, and 4) monitoring.

ML applications are highly data-driven, and another field that is also highly
data-driven is the database field.
Wang et. al review challenges and opportunities with combining the fields of
DL and databases~\cite{wang2016database, re2015machine}.
As the database community has rich experience working with system optimization,
the DL community could benefit from taking advantage of database
techniques to accelerate training speed.
In particular, the fields of distributed computing and memory management are
central to both the database and DL
community~\cite{dean2012large,jia2014caffe,abadi2016tensorflow}.

Another field closely related to ML and DL is that of big data.
Chen et. al review challenges and perspectives in the intersection
of these two fields~\cite{chen2014big}.
The area of big data offers opportunities but also adds significant engineering
challenges.

\section{Conclusions}
\label{sec:conclusions}

As we have seen in recent years, companies such as Google, Microsoft, and
Facebook are transforming their companies to make AI, and specifically DL,
become a natural component of their products across the
line~\cite{parekh2017designing,zerega2017ai}.
However, there are many challenges with building production-ready systems
with DL components, especially if the company does not have a large
research group and a highly developed supporting infrastructure.

The main goal of this research is to identify and outline main SE challenges
with building systems with DL components.
Seven projects were described to exemplify the potential for making use of
the ML and specifically the DL technology.
For these projects, the main problematic areas and challenges with building
these systems were identified.
To clarify these problematic areas in more detail, a set of
12 challenges were identified and described in the areas of: development,
production, and organizational challenges.
The main focus of this paper is not to provide solutions, but rather to outline
problem areas and, in that way, help guide future research.
In addition, the outlined challenges also provide guidance on
potential problem areas for companies interested in building high-quality
DL systems.

One clear conclusion of this work is that, although the DL
technology has achieved very promising results, there is still a significant
need for further research into and development in how to easily and efficiently
build high-quality production-ready DL systems.
Traditional SE has high-quality tools and practices for reviewing, writing
test, and debugging code.
However, they are rarely sufficient for building production-ready systems
containing DL components.
If the SE community, together with the DL community, could make an effort
in finding solutions to these challenges, the power of the DL technology
could not only be made available to researchers and large technology
companies, but also to the vast majority of companies around the world.

\IEEEtriggeratref{50}
\bibliographystyle{IEEEtran}
\bibliography{IEEEabrv,article}

\end{document}